\documentclass[12pt,english]{article}

\font\Bbb = msbm10

\def\FF{{\cal F}}

\def\DD{{\cal D}}

\def\BC{\mbox{\Bbb C}}
\def\BD{\mbox{\Bbb D}}
\def\BZ{\mbox{\Bbb Z}}
\def\BR{\mbox{\Bbb R}}

\def\be{\begin{equation}}
\def\ee{\end{equation}}
\def\bea{\begin{eqnarray}}
\def\eea{\end{eqnarray}}
\def\f{\frac}

\def\La{\Lambda}
\def\al{\alpha}

\def\th{\theta}

\newcommand{\QED}{\mbox{\rule[-1.5pt]{6pt}{10pt}}}

\newtheorem{defi}{Definition}[section]
\newtheorem{prop}{Proposition}[section]
\title{{\bf The Thom Class and Localization of SUSY QM Generating Functional}}
\author{V\'a\v{n}a O.\footnote{
Mathematical Institute
of Charles University, vana@stones.com }}
\date{December 11, 2000}

\begin{document}
\maketitle
\begin{abstract}
We demonstrate the usage of explicit form of the Thom class found by Mathai and Quillen for
the definition of generating functional of a simple supersymmetric quantum
mechanical model.
\end{abstract}

\section{Backgrounds on superformalism}
\begin{defi}
The Grassmann algebra $\Lambda_n(\BC)$ is an $\BC$ algebra freely generated by the set of
$n$ anticommuting generators $\{\eta_1,\dots,\eta_n\}$, i.e.

\be
\{\eta_i,\eta_j\}=\eta_i\eta_j+\eta_j\eta_i=0\ \forall i,j\in \hat{n}
\ee
Similarly we define $\La_n(\BR)$.
\end{defi} 
The straightforward generalization of this is the notion of super$(\BZ_2)$ graded
algebra.

\begin{defi}
Let $M$ be a $\BZ_2$ graded module over a $\BZ_2$ graded ring R. Moreover, let $M$ be endowed with a grading
compatible multiplication
\be
m:M\otimes_R M\rightarrow M
\ee
Then $M$ is called the superalgebra over R.
\end{defi}
Clearly, the Grassmann algebra $\Lambda_n(\BC)$ is an example of superalgebra over $R=\BC$. It's the linear
span of $1$ and the monomials 
\be
\label{mon}
\{\eta_{i_1}\dots\eta_{i_k}\mid 1\le i_1<\dots<i_k\le n,\ k\in\hat{n}\}
\ee 
$R$ is considered with trivial grading here and an element in $\Lambda_n(\BC)$ is even (odd) if it has in the
expansion with respect to (\ref{mon}) the even (odd) monomials only.

In a superalgebra one can define so called {\it generalized Lie bracket} putting
\be
\label{bra}
[a,b]:=a\cdot b -(-1)^{deg(a)deg(b)}b\cdot a
\ee
for homogeneous elements and expanding it by linearity. A superalgebra is called {\it commutative} if the
bracket (\ref{bra}) vanishes. Given two superalgebras $A$ and $B$ one can form
the tensor product $A\otimes B$ of them in the usual way, i.e. one makes the tensor product of
underlying $\BZ_2$ graded modules and endows it with the multiplication
defined by $(a\otimes b)\cdot(a'\otimes
b'):=(-1)^{deg(b)deg(a')}aa'\otimes bb'$. 

Suppose we have an arbitrary superalgebra. Then one can define the Pfaffian by 
the following

\begin{defi}
Suppose $n=2m$ and let $\omega$ be a skew-symmetric $n\times n$ matrix of even elements in an arbitrary superalgebra. The
Pfaffian of $\omega$ with respect to the vector of odd elements
$\{\eta_i\}_{i=1}^n$ is defined as 
\be
\frac{1}{m!}\Bigl(\frac{1}{2}\eta^t\omega\eta\Bigr)^m=Pf(\omega)\eta_1\dots\eta_n
\ee
where $(\cdot)^t$ denotes the transpose.
\end{defi}
Another useful tool in calculations with superquantities is {\it Berezin} or
{\it fermionic} integral.

\begin{defi}
The Berezin integral of an element $x$ in $\La_n(\BC)$, resp. $\La_n(\BR)$ generated by
$\{\eta_1,\dots,\eta_n\}$ is defined as the coefficient of $x$ staying before
the $\eta_1\dots\eta_n$ monomial, i.e. when  

\be
x=\sum_{k=0}^n\sum_{1\le i_1<\dots<i_k\le n}\al^{(k)}_{i_1\dots i_k
}\eta_{i_1}\dots\eta_{i_k}
\ee
then $\int{\cal D}\eta\,x:=\al^{(n)}_{1\dots n}$. One can extend this
definition to the elements of $A\otimes \La_n(\BC)$ where $A$ is an arbitrary
superalgebra as $\int{\cal
D}\eta\, a\otimes x:=a\int{\cal D}\eta\,x$. 
\end{defi} 
 
With the notion of Berezin integral one can express the Pfaffians 
as

\be
Pf(\omega)=\int\DD\eta\,e^{\frac{1}{2}\eta^t\omega\eta}
\ee
and the following holds

\begin{prop}\label{bas}
Let $A$ be any commutative superalgebra, $n=2m$, $\omega$ a skew-symmetric $n\times n$ matrix of even
elements, $\{\eta_i\}_{i=1}^n$ be a vector of odd elements. Then
\be\label{pfaf}
e^{\frac{1}{2}\eta^t\omega\eta}=\sum_{I}Pf(\omega_{I})\eta^I
\ee
where $I$ runs over all subsets of $\hat{n}$ with even cardinality and for
$I=\{i_1,\dots,i_k\}$
, $i_1<\dots <i_k$, $\eta^I$ denotes $\eta_{i_1}\dots\eta_{i_k}$ while
$(\omega_I)_{i'j'}$ denotes the submatrix of $(\omega)_{ij}$ with $i'$, $j'$ in $I$. 
\end{prop}

\vskip 1ex
\noindent
{\bf Proof}. Suppose for the moment $A$ is the Grassmann algebra generated by
$\{\eta_1,\dots,\eta_n\}$. Consider the subalgebra generated by $\eta_i$, $i\in
I$. Then the homomorphisms $h_I$ which kill all $\eta_k$,
$k\notin I$ sends $e^{\frac{1}{2}\eta^t\omega\eta}$ to the corresponding
Gaussian expression constructed from $\omega_I$. Hence 

\be
e^{\frac{1}{2}\eta^t\omega\eta}=\sum_I\int\DD\eta^I\,
h_I\Bigl(e^{\frac{1}{2}\eta^t\omega\eta}\Bigr)=\sum_I\int\DD\eta^I\,e^{\frac{1}{2}(\eta^I)^t\omega_I\eta^I}=\sum_I
Pf(\omega_I)\eta^I
\ee
which verifies (\ref{pfaf}) in this case. Suppose now $A$ is any commutative superalgebra. Both elements in (\ref{pfaf})
can be interpreted as an elements of free commutative superalgebra
$S\otimes\La_n(\BC)[\eta_1,\dots,\eta_n]$, where $S$ is the algebra of
polynomials in variables $\omega_{ij}$. These elements are equal since
polynomials are determined by it values. Therefore we can apply realization
homomorphism which takes the elements $\omega_{ij}$, $\eta_i$ of the former to
corresponding element of $A$. Thus (\ref{pfaf}) holds in $A$.
\hspace*{\fill}\QED

Using this proposition one can state

\begin{prop}\label{ber}
Let $A$ be any commutative superalgebra, $n=2m$, $\omega$ a skew-symmetric $n\times n$ matrix of even
elements, $\{J_i\}_{i=1}^n$ be a vector of odd elements. Then

\be
\int\DD\eta\,
e^{\frac{1}{2}\eta^t\omega\eta+J^t\eta}=\sum_{I\ even}\epsilon(I,I')(-1)^{\frac{1}{2}|I'|}Pf(\omega_I)J^{I'}
\ee
where $I'$ denotes the complement of $I$ in $\hat{n}$ and $\epsilon(I,I')$ is
defined as $\eta^I\eta^{I'}=\epsilon(I,I')\eta_1\dots\eta_n$.
\end{prop}

\vskip 1ex
\noindent
{\bf Proof}. uses \ref{bas}. See \cite{math1} for further details.
\hspace*{\fill}\QED

\section{Thom class construction and Euler \\
characteristic}

Suppose now $n=2m$ and we have a compact manifold $M$, $dim M=n$ with spin
structure,  i.e. we have an isomorphism of $TM$ with bundle $P\times_{Spin(n)}
V$ associated to principal spin frame bundle $P$ over $M$ and standard
representation of $Spin(n)$ on $V=\BR^n$. Hence $P\times V$ is a principal
$Spin(n)$ bundle over $TM$ which is equal to $\pi^*P$, $\pi$ is the $TM$
projection. The forms $\Omega(TM)$ can be identified with basic forms of
$\Omega(P\times V)$, i.e. the forms which satisfies

\bea
R^*_g\omega=\omega\\
i_X\omega=0
\eea
for all $g$ in $Spin(n)$ and all $X$ in its Lie algebra. In particular, the
identification isomorphism is given by pull-back with respect to the projection
$f:P\times V\rightarrow TM$. 

Suppose now we have the spin connection $\th$ in $P$. This connection may be pulled
back to $P\times V$. We denote this connection by the same symbol. 
If one consider its curvature

\be
\Omega=d\th+\th\wedge\th
\ee
then one can define the form on $\Omega(P\times V)_{basic}\simeq\Omega(TM)$
putting

\be\label{U}
U=\pi^{-m}e^{-x^2}\sum_{I\
even}\epsilon(I,I')Pf(\f{1}{2}\Omega_I)(dx+\th x)^{I'}
\ee
In fact, one can show that this form is closed (see \cite{math1}) and
integrates to $1$ over the fibres as is easily seen from the identity

\be
\pi^{-m}\int_{\BR^n}e^{-x^2}\,d^n x=1
\ee

We recall the Thom class definition. Let $\FF$ be an oriented vector bundle
with scalar product and
let $\BD\FF$ denotes the unit disk bundle of $\FF$. Then one has the Thom isomorphism

\be
\pi_*:H^i(\FF,\FF\setminus\BD\FF)\rightarrow H^{i-n}(M)
\ee

\begin{defi}
The Thom class is an element of $H^n(\FF,\FF\setminus\BD\FF)$ defined as

\be
u(\FF):=\pi_*^{-1}(1)
\ee
\end{defi}
When passing to the form representation of cohomology classes $\pi_*$
corresponds to integration over the fibres and thus the Thom class is
represented by a closed form with support in $\BD\FF$ which integrates to $1$
over the fibres.
  
The form (\ref{U}) represents the Thom class of $TM$ in the following sense.
This form doesn't have the support in $\BD TM$ but one can define a fibrewise
diffeomorphism, namely 

\bea
f:TM\rightarrow\BD TM\\
\nonumber
f(x):=\f{x}{\sqrt{(1+\|x\|^2)}} 
\eea
and define the complex $\Omega_{d}(TM)$ as
$f^*\Bigl(\Omega(TM,TM\setminus\BD TM)\Bigr)$. Then since integration is
invariant under orientation preserving diffeomorphisms we can extend $\pi^*$ as

\be
\tilde{\pi^*}:\Omega_d(TM)\rightarrow\Omega(M)
\ee
which induces the isomorphism on cohomology. Thus $U\in\Omega_d(TM)$
represents the Thom class since it's closed and integrates to $1$ over the
fibres. 

Now we turn back to
proposition \ref{ber} and apply it for the case of $A:=\Omega(P\times V)$ and
$J=dx+\th x$. Thus the form $U$ may be expressed as

\be\label{U1}
U=\pi^{-m}\int\DD\eta\,e^{-x^2+\f{1}{2}\eta^t\Omega\eta+(dx+\th x)^t\eta}
\ee
If we denote by $s:M\rightarrow TM$ section with isolated zeroes then $s^*U$ represents
Euler class of $M$ and we have the well-known formula

\be\label{h}
\int_M s^*U=\sum_i {\rm Ind}_i(s)
\ee
This can be proved in
greater generality for an arbitrary oriented vector bundle over a compact
manifold, considering the family of sections $\{s_t\}_{t\in\BR^+}$, $s_t:=ts$
for such $s$. All the
forms $s_t^*U$ represents the same cohomology class and thus for $t\rightarrow
0$ we obtain the Euler class by the definition. The integral
$\int s_t^*U$ remains the same for all $t$ and 
 for $t\to+\infty$ one obtains RHS of (\ref{h}). However, 
in the case of $TM$ famous Hopf theorem identifies RHS of (\ref{h}) with
Euler characteristic $\chi(M)$. 
	
Finally if we choose such a section $s$ then we have from (\ref{U1})

\be\label{eul}
\chi(M)=\pi^{-m}\int_M\int\DD\eta\,e^{-s^2+\f{1}{2}\eta^t
(s^*\Omega)\eta+(\nabla s)^t\eta}
\ee
which works even for $s=0$ from the reasoning above.

\section{A physical interpretation: SUSY quantum mechanics}
We are going to show that (\ref{eul}) is in fact the generating functional for
SUSY quantum mechanics. In physics, however, one has to start with the loop
space in order to obtain the action in integral form.
This is defined as follows

\begin{defi}
Let $M$ be a differentiable manifold. Loop space $LM$ of $M$ is defined to be
the set of all smooth mappings from $S^1$ to $M$.
\end{defi}
Of course, one has to define the topology and differentiable structure on $LM$
(see \cite{seg}), but we can treat it at least formaly in physics. The tangent
bundle of $LM$ is defined as 

\be
T_xLM:=\{X\in LTM\mid \pi_{TM}X(t)=x(t)\ \forall t\in S^1\}
\ee
If $M$ is equipped with Riemannian
metric then the induced metric on $TLM$ is defined as

\be
\hat{g}_x(V_1(x),V_2(x)):=\int_{S^1}dt\,g_{x(t)}(V_1(x)(t),V_2(x)(t))
\ee
where $x\in LM$.

Now if we put $dx^\mu=\psi^\mu$ and $s=\dot{x}^\mu$, i.e. we identify fermionic fields with
coordinates in $T^*LM$ and choose our section to be the 'time' derivative of
bosonic fields, then we can, at least formaly, identify the exponent in
(\ref{eul}) with the action of SUSY QM which in coordinates looks like 

\be
S_{QM}=\int_{S^1}dt\,\Bigr[-g_{\mu\nu}\dot{x}^\mu\dot{x}^\nu+\f{1}{2}R^{\mu\nu}_{\rho\sigma}\bar{\psi}_\mu\psi^\rho\bar{\psi}_\nu\psi^\sigma+\bar{\psi}_\mu\nabla_t\psi^\mu\Bigr]
\ee
Here $\nabla_t$ is the 'covariant derivative' defined as

\be
\nabla_t\psi^\mu(t):=\dot{\psi}^\mu(t)+\Gamma^\mu_{\rho\sigma}\dot{x}^\rho\psi^\sigma
\ee
where $\Gamma^\mu_{\rho\sigma}$ are Christoffel symbols of Levi-Civita
connection induced by $g$. The term $\nabla s$ in (\ref{eul}) corresponds to

\be
\nabla_t\psi^\mu\leftrightarrow\nabla s
\ee

Finally, we consider the generating functional of this physical model

\be
Z_{S_{QM}}=\int_{LM}\DD\psi\DD\bar{\psi}\DD x\,e^{iS_{QM}[\psi,\bar{\psi},x]}
\ee
This integral over loop space is not correctly defined, however, we can apply
formaly on it (\ref{h}) and 'localize' it at zeroes of $s=\dot{x}^\mu$. But
this means to integrate over the space of constant loops, i.e. the manifold
$M$. This means we recover the (\ref{eul}) for $s=0$ in this case and thus 'regularized
Euler characteristic' can be identified with Euler characteristic of $M$.
This is the argument physicests use proving the fact that SUSY QM is an example
of topological field theory.

\end{document}